# Prehistory of Transit Searches

Danielle BRIOT[1] & Jean SCHNEIDER[2]

1) GEPI, UMR 8111, Observatoire de Paris, 61 avenue de l'Observatoire, F-75014, Paris, France
danielle.briot@obspm.fr
2) LUTh, UMR 8102, Observatoire de Paris, 5 place Jules Janssen, F-92195 Meudon Cedex, France
jean.schneider@obspm.fr

## Abstract

Nowadays the more powerful method to detect extrasolar planets is the transit method, that is to say observations of the stellar luminosity regularly decreasing when the planet is transiting the star. We review the planet transits which were anticipated, searched, and the first ones which were observed all through history.
Indeed transits of planets in front of their star were first investigated and studied in the solar system, concerning the star Sun. The first observations of sunspots were sometimes mistaken for transits of unknown planets. The first scientific observation and study of a transit in the solar system was the observation of Mercury transit by Pierre Gassendi in 1631. Because observations of Venus transits could give a way to determine the distance Sun-Earth, transits of Venus were overwhelmingly observed. Some objects which actually do not exist were searched by their hypothetical transits on the Sun, as some examples a Venus satellite and an infra-mercurial planet. We evoke the possibly first use of the hypothesis of an exoplanet transit to explain some periodic variations of the luminosity of a star, namely the star Algol, during the eighteen century. Then we review the predictions of detection of exoplanets by their transits, those predictions being sometimes ancient, and made by astronomers as well as popular science writers. However, these very interesting predictions were never published in peer-reviewed journals specialized in astronomical discoveries and results.
A possible transit of the planet β Pic b was observed in 1981. Shall we see another transit expected for the same planet during 2018 ?
Nowadays, some studies of transits which are connected to hypothetical extraterrestrial civilisations are published in astronomical peer reviewed journals. So we can note that the discovery of exoplanets is being modified the research methods of astronomers. Some studies which would be classified not long ago as science fiction are now considered as scientific ones.

## 1. Introduction

The discovery of the planet orbiting the star 51 Pegasi, now planet named 51 Peg b, has been a striking discovery for all the astronomical community and even more for the mankind. Actually this discovery was expected and hoped for a very long time.
Various methods exist for the detection of exoplanets and some of them were anticipated for a long time. A good synthetic presentation of all these methods is given by a figure of Michael Perryman (Perryman 2000). This figure is known as *The Perryman tree* because the display of the different methods is organized into a hierarchy. The first efficient method was the detection by observations of small periodic variations of radial velocities. During several years, it was the only successful method. Nowadays the most efficient method is the detection of transits of the planets in front of their star, implying a periodic decrease of the star's luminosity. This method is essential for determining some physical parameters of the planet as period, diameter, mass, chemical atmospheric composition. The present study is mostly dedicated to preliminary studies predicting, looking for or observing transits. Before that, we briefly review some methods for planet detection and their precursory studies. The search for other worlds have been existing since antique ages. Nowadays this formulation is interpreted as the search for planets around other stars than the Sun. However the meaning of other worlds has then been totally different than this representation. Two fundamental astronomical discoveries have been necessary so that the words `*search for other worlds*' correspond to the search for extrasolar planets. The first discovery is the heliocentric system established by Nicolaus Copernicus (1473-1543) in the book *De revolutionibus orbium coelestium*, published in 1543, that is to say the planets are orbiting the Sun, and so Earth is no more the centre of the world (Copernicus 1543). The second discovery is the understanding that stars are other Suns. During the seventeenth century, many unsuccessful searchs for determination of stellar distances have implied that stars are much more remote than it was supposed. Then stars are intrinsically very luminous, as is our Sun. Because stars are objects similar to the Sun, it is highly likely that they are surrounded by a planetary system. As soon as 1686, Bernard le Bouyer de Fontenelle (1657-1757) wrote in his the book *Entretiens sur la pluralité des mondes* i.e. *A conversation on the Plurality of Worlds*: `*Every fixed star is a sun, which diffuses lights to its surrounding worlds*' (Fontenelle1686). This book was a best-seller, it was re-edited many times and translated in many languages. Its influence throughout the occidental world was very important. So, as soon as the second part of the seventeenth century the existence of extrasolar planets was considered. The discovery of these planets was made in 1995, more than three centuries later (Mayor and Queloz 1995).

This paper is devoted to some studies in the past which searched for and predicted methods ahead of their time for the detection of possible planets around other stars than the Sun, then from the seventeenth century. After a rapid review of some precursory studies of various methods, we will focus specially on the studies predicting some planetary transits and establishing detection methods for them, first in solar system then outside solar system.

**2. Predictory studies of various methods for exoplanet detection**

1. Imaging - After centuries of philosophical speculations the first scientific approach to the detection of exoplanets was due to Christiaan Huygens (1629-1695) as soon as 1698, by imaging (Huygens1698). In the book *Kosmotheoros*, Huygens at once admitted that no planet could be seen: `*For let us fancy our selves placed at an equal distance from the Sun and fixed Stars; we would then perceive no difference between them. For, as for all the Planets that we know see attend the Sun, we should not have the least glimpse of them, either that their Light would be too weak to affect us, or that the Orbs in which they move would make up one lucid point with the Sun*' (Huygens 1698).

2. Astrometry - For example Kaj Aage Strand (1907-2000) wrote in 1943 about an unseen companion in the double star system 61 Cygni: *With a mass considerably smaller than the smallest known stellar mass, the dark companion must have an intrinsic luminosity so extremely low that we may consider it a planet rather than a star. Thus planetary motion has been found outside the solar system...*' (Strand 1943).

3. Radial velocity - This method was predicted by Otto Struve (1897-1963) in 1952: `*A planet ten times the mass of Jupiter would be easy to detect, since it would cause the observed radial velocity of the star to oscillate with ± 2 km s$^{-1}$*' (Struve1952). As we know, this method was very successful to detect the first exoplanet and many other ones. It was the only method efficiently used during several years.

4. Multiplanet perturbations - This method was successfully used in the Solar System to discover the Neptune planet by Urbain Le Verrier (1811-1877) in 1846 (Le Verrier 1846).

## 3. Transits in the solar system

The history of astronomy mentions several observations of supposed transits on the Sun anterior to the first observations with an optical instrument. The question is: was it real transits or more probably sunspots ?

### 3.1. Fictional transit : solar spots, patches or planets

As soon as sunspots are observed, two hypotheses have been put forward to explain their origins : patches on the Sun or transit in front of the Sun of unknown infra-mercurial planets.
In 1613, Galileo Galilei (1564-1642) announced that he discovered and observed some spots in front of the Sun. At once, a controversy appeared about some anterior observations of sunspots by Thomas Harriot (1560-1621), Christoph Scheiner (1575-1650), David Fabricius (1564-1617) and his son Johannes Fabricius (1586-1615). Furthermore, historians have made inventories of sunspots observations long ago, in various civilisations, with naked eyes or with a camera obscura. An example of these inventories can be found in Vaquero (Vaquero 2007).
Jean Tarde (1561-1636) was a canon in Sarlat, in the Perigord (South-West of France). He went to visit Galileo in 1614. He observed and studied sunspots during four years. His observations are probably the most or among the most extendend period observations of sunspots at this epoch. He carried out his observations with a scientific method. He interpreted the sunspots as small planets passing between Mercury and the Sun. He named the planets that he supposed he observed *Borbonia sidera*, i.e. *Bourbonian planets*, from the dynasty name of the king of France, to honour Louis XIII, the king of France, as Galileo named Medician planets the four Jupiter satellites that he discovered to honour the Medicis princes. He published a book in latin *Borbona sidera* in 1620, translated in French in 1622, *Les astres de Borbon* (Tarde 1620, 1622). We have to emphasize that he used the Copernic system, i.e. the Earth orbiting the Sun, whereas he was a priest of the Catholic Church. He carried on his observations with a great perseverance. He noted that those planets move with different velocities, and are moving slowly that Mercury. The third law found by Johannes Kepler (1571-1630) establishing the relation between the period of a planet and its distance to the Sun was published only in 1619 (Kepler 1619), and probably Jean Tarde did not know it when he wrote his book published in 1620. As many scientists of this epoch, he used a religious argument to refute the theory of sunspots belonging to the Sun. He wrote that spots on the Sun are impossible because God choose the Sun as place of residence : `*In sole posuit tabernaculum Suum*.'. The place chosen by God to stay could not be corrupted. Tarde quoted this sentence in Latin even in the french version of his book. He did not indicate the origin of it, that means that this sentence was being very known by everypeople. Actually this sentence is a part of the psalm 19, in the Bible version called the Vulgate, translated from the Hebrew to Latin by St Jerome at the end of the 4th century AD. This very popular version of the Bible was used by the Catholic Church during many centuries and it was the first book ever printed by Gutenberg in 1455. However this sentence, which is used by Tarde as a basis for his argumentation denying that the origin of dark patches seen on the Sun are sunspots, results from a mistake in the translation from Hebrew, or in a copy of the original translation. The real meaning is something like : `*In the heavens God has pitched a tent for the sun*', as it can be seen in any other translation, in many various languages. However, the idea that the Sun is pure and cannot be corrupted nor soiled corresponds to the description of the world according to the Aristote's philosophy. Some more information about life and work of Jean Tarde can be found in Baumgartner (Baumgartner 1987).

### 3.2. Transit of Mercury : the first transit really observed

In 1627, Johannes Kepler (1571-1630) published *Tabulae Rudolphinae*, the *Rudolphine Tables* (Kepler 1627) so called in honour of the former emperor Rudolph II of Habsburg (1552-1612). These astronomical tables are based on the three laws concerning the planet motions published by Kepler in 1609 and 1618 and using the observations of Tycho Brahé (1552-1601). The discovery of logarithms by the Scot John Napier (1550-1617) in 1614 (Napier 1614) was greatly appreciated by Kepler and give facilities for the making of the Tables. In 1630 he published ephemerides for the years 1629 to 1639, based on his *Rudolphine Tables* in which he included a *Reminder for Astronomers and people studing celestious objects* (*Admonitio ad astronomos, rerumqve coelestium studiosos* that we will name simply *Admonitio*) (Kepler 1629) indicating that it will be a transit of Venus in front of the Sun the 6th of December 1631, visible from America, according to his calculations. However as a little mistake could possibly exist in his predictions, he advice European astronomers to observe the Sun during this day. Moreover on the 7th of November of the same year it will happen a transit of Mercury in front of the Sun. However, because of difficult Mercury observations, this is no certainty in this date and Kepler adviced to European astronomers to observe from the 6th and continue observations up to the 8th of November. *Admonitio* is distributed through Europa to educated people.

Using Kepler's *Admonitio* the French scientist Pierre Gassendi (1592-1655) very carefully prepared the observation of the Mercury transit. He was in Paris during the planned days. Kepler has adviced to project image of the Sun on a paper with a refracting telescope, or with a simple *camera obscura* in the absence of any telescope. A camera obscura can be equiped with an optical instrument, for example a single lens or a refracting telescope, as well as a mirror to straighten the images obtained. Gassendi owned already an instrumentation that he used to observe sunspots and solar eclipses. In the camera obscura that he used, luminous rays coming from the Sun passed through a Galilean telescope and formed a Sun image on a sheet of paper. The adjustment was made as this image diameter would be around twenty-five centimeters. He draw a similar circle which he divided the diameter in sixty equal parts. In another room immediately below an assistant using a two-foot quarter-circle instrument was to observe and note the height of the Sun when Gassendi indicated stamping his foot. So it is obvious that this observation has been prepared in a really scientific way. The weather was in part cloudy and it was impossible to observe the Sun before the 7th of November. On this morning Gassendi observed a very little black patch that he supposed in first to be a sunspot, because he was surprised by the smallness of this patch, and very soon he realized that for the first time he observed a planet, that is Mercury, in front of the Sun. Gassendi described in detail his observations in a book entitled *Mercurius in sole visu et Venus invisa (Mercure visible on the Sun and Venus invisible)*, published en 1632 (Gassendi 1632).

However for several reasons very few people could observe this first Mercury transit. One of the reasons was the rainy or cloudy weather in these days, which is very frequent in November in a great part of Europe. Another reason was the unexpected smallness of Mercury, so observers who used a camera obscura without any optical instrument could not see Mercury. The transit of Mercury was also observed by Johan-Baptist Cysatus, the former pupil of Christoph Scheiner, in Innsbruck (Austria), by Johannes Remus Quietanus, physician and mathematician of the Emperor Mathias in Rouffach (Alsace) and by an anonymous Jesuit in Ingolstadt (Bavaria). We do not know circumstances of any of these observations, so Gassendi's observation is the only one which it is possible to deduce some
astronomical conclusions and so the only one which can be considered as
really scientific.

A transit of Venus was supposed observed in other circumstances. In 1607,
from Tycho Brahe's observations, Kepler calculated that a transit of
Mercury in front of the Sun will happen at the end of May. He observed the
Sun on the 28th of May with a makeshift camera obscura, without any lenses.
Actually, he detected a black spot on the Sun that he supposed to be
Mercury. However, when sunspots have been discovered by observations
carried on with optical instruments, Kepler understood then that he
observed some sunspots.

The observation in 1631 by Gassendi of the Mercury transit is of a real
importance. It allowed astronomers to correct Kepler's data about Mercury.
The Mercury inclination on the ecliptic plane and the trajectory of
Mercury became much more precise. A very important result was a new
estimation of the Mercury diameter, much smaller that though up to this
time. This last point implied that the planet diameter could not be deduced
from the luminosity, or from the telescope observations. Some information
about the importance of the first observation of a planet transit can be
found for example in Van Helden (Van Helden 1876).

### 3.3 Some information about Venus transits

In his *Adminitio*, Kepler announced a transit of Venus in front of the Sun
predicted for the 7th of December of 1631, that is one month after the
Mercury transit. However this transit could not be observed from Europe,
what explains the second part of the title of the book written by Gassendi:
*et Venus invisa*. Kepler expected that the next Venus transit will happen in
1761, that is to say 120 years later. So the astronomers which were
interested directed their studies to other subjects. However, in England,
Jeremiah Horrocks (1618-1641) studied the *Rudolphine Tables* of Kepler and
he determined in October 1639 that the next transit will actually happen
the 4th of December 1639, according the Gregorian calendar, that is only 8
years after the transit of 1631. He notified his friend and correspondant
William Crabtree (1610-1644) in order that he would observe this
phenomenon. They were the only first observers of a Venus transit. Nowadays
we know that the Venus transits happen 4 times during a cycle of 243 years,
the intervals between the transits being 8 years, 105.5 years, 8 years and
121.5 years. The Mercury transits are much more frequent because they
happen 13 or 14 times during a century.

After having observed a Mercury transit in 1677, Edmund Halley (1656-1742)
showed that observations of the transits of the inferior planets, Mercury
or Venus, from places of different latitudes on Earth, allow to determine
the distance Sun-Earth and then all the distances in the solar system. The
Venus transits are more easy to observe than the Mercury transits. Using
the Halley's method, it would be necessary to precisely determine the
times of the contacts between the planet Venus and the limit of the Sun
surface, observed from different places on Earth. The following Venus
transits happened in 1761 and 1769, then in 1874 and 1882. To observe these
Venus transits many perilous expeditions were launched worldwide from more
and more countries. The relations of these expeditions represent a very
interesting part of the history of astronomy, often picturesque and
sometimes tragic.

Unfortunately the results were not as good as hoped, because the observers
faced the black drop phenomenon. They had to determine the precise times of
the 'second contact' and the 'third' contact. The second contact
corresponds to the moment when the surface of Venus appears completely on
the Sun surface, the edge of Venus being tangential to the edge of the Sun.
The third contact corresponds to the moment when the edge of Venus is
tangential to the edge of the Sun, the surface of Venus totally appearing on the surface of the Sun, just before the last phase of the transit. However, the moment of the second contact was quite impossible to precisely determine because the black disc of Venus seemed to remain linked to the edge of the Sun by a dark 'neck'. The surface of Venus did no more appear as circular, but almost pear-shaped. The problem existed as well for the determination of the precise moment of the third contact. The accuracy of the time determinations of the second and the third contacts was hoped within about a second by Halley. Due to the black drop effect, the accuracy of timing became like a minute. When due to the instrumental progress, astronomers understood how to eliminate the black drop phenomenon, other more precise ways to determine the distances in the solar system were discovered. Nowadays the more accurate method use some laser.

### 3.4. Research of transits of objects which actually do not exist

The history of searchs for transits in the solar system is not always successful.
Numerous observations have been carried out to detect a satellite for the Venus planet as well as some infra-Mercurial planets. A list of *Observations or supposed observations of the Transits of Infra-Mercurial Planets or other Bodies across the Sun's Disk from 1761 to 1865* is displayed by Ledger (1879). The precise trajectory of the planet Mercury, and particularly the advance of its perihelion – the point on its orbit when Mecury is closest to the Sun- cannot been explained using only the Newtonian theory. As the French astronomer Le Verrier discovered in 1846 the planet Neptune by calculations from the trajectories of other planets, he tentatively explained the trajectory of Mercury by an hypothetical planet orbiting between Mercury and the Sun. This infra-mercurial hypothetical planet named Vulcain has been researched for a long time, by astronomers and amateurs observers and sometimes has been believed to be observed. The solution for this problem has been obtained only in 1915 when Albert Einstein discovered the General Theory of Relativity. Actually, the corrections to Newton's theory due to the General Theory of Relativity explains the advance of the perihelion of Mercury. Obsviously Le Verrier could not know this theory.

### 3.5. Other occultations or transits in the solar system

Let us briefly recall the importance of scientific discoveries by other transits or occultations in the solar system. The word occultation is used when the transiting object hides the whole transited object or a large part of it. The regular movments of the Galilean satellites of Jupiter, hidden or shadowed by the planet allowed the discovery of the velocity of light during the 17th century.

## 4. Early predictions of detection of exoplanets by the transit method

Thanks to the Kepler space telescope, the detection of extrasolar planets by observing the periodical decreasing of the luminosity of the star due to the transit of the planets between the star and the observer is now the most efficient way to detect extrasolar planets. Before this method has been used successfully, this way of detection was announced in some very premonitory studies, and sometimes a long time ago before the first transit observed.

### 4.1 Some hypothesis for explaining Algol variations

Algol is a regularly variable star. The first observations of these variations are generally attributed to Geminiano Montanari (1633-1687) from 1668 to 1677, and Giovanni Philippo Maraldi (1665-1729) around 1693 and 1694. However these variations are probably known for a very long time because the origin of the name of the star is generally considered as arabic and meaning "The Demon". This star, as well many other variable stars, was intensively observed during the eighteenth century by two close friends and collaborators, Edward Pigott (1753-1825) and John Goodricke (1764-1786). Goodricke determined the period of variations as 2 days, 20 h, 45 min. It is remarquable that this differs only a few minutes from the modern value (Goodricke 1783).

These very carefull observers made some assumptions about the cause of these regular variations. Goodricke wrote: *'If it were not perharps too early to hazard even a conjecture on the cause of this variation, I should imagine it could hardly be accounted for otherwise than either by the interposition of a large body revolving round Algol, or some kind of motion of its own, whereby part of its body, covered with spots or such like matter, is periodically turned towards the earth. But the intention of this paper is to communicate facts, not conjectures...'*(Goodricke 1783). However, Michael Hoskin having had the opportunity to study the journal of Piggot, attributes to him the hypothesis of a transit of a large body, planet or satellite (Hoskin 1979). Some years latter, Piggot wrote *'Hitherto the opinion of astronomers concerning the changes of Algol's light seem to be very unsettled ; at least none are universally adopted, though various are the hypotheses to account for it ; such, as supposing the star of some other than a spherical form, or a large body revolving round it, or with several dark spots or small bright ones on its surface, also giving an inclination to its axis, &c. ...'*(Piggot 1785). Let us notice that this argumentation was regarded interesting enough to be published again in a French journal *Journal Encyclopédique*. This is possibly the first use, and at least one the the first ones, of the hypothesis of a planet transit to explain regular variations of a star light, although the existence of planets orbiting around stars was considered as plausible for more than a century (see for example Fontenelle 1683). Nowadays we know that Algol is a semi-detached binary star with a mass transfer from one component to the other one, at present the less massive star being the more evolved. The name Algol is now used for all the stars having the same evolutionary path.

   4.2 Predictive announcements of discoveries of  exoplanets by transits

The first known indication of a possible detection of exoplanets by transit was predicted by Dionysius Lardner (1793-1859) in a book of popular science. He studied the periodic variable stars and listed all the hypotheses proposed to explain the phenomena. The fifth hypothesis is : `*It has been suggested that the periodical obscuration or total disappearance of the star may arise from the transits of the star by its attendant planets*' (Lardner 1853). Lardner was a Irish populariser of science. David Belorizky (1901-1982) was an astronomer at the Marseille observatory. In 1938 he wrote a paper about variations of the Sun in *L'Astronomie*, a magazine for amateurs-astronomers (Belorizky 1938). In a paper intitled "The Sun, as variable star" he studies the different ways for discovering other planetary systems. He explains that the variation of the radial velocity of the sun due to the presence of Jupiter could not be detected by the spectrographs existing at this time. He completely rejects the possibility to observe a planet orbiting a star, neither by eye, nor photographically, considering the magnitude of a planet located at a stellar distance and considering also the very large ratio between the luminosity of the star and the luminosity of the planet. He studied the variation of the luminosity of the Sun due to some transits of Jupiter observed from a other planetary system and wrote : `*The only way that we* *see at the moment to possibly detect existence of planets in other worlds is the photometry with a precision of 1/100 magnitude, which is the precision of current photo-cells*'. The life of David Belorizky shows some interesting coincidences. He was born in Russia and emigrated to France during the Twenties. To escape the Jew extermination during the World War Two and the Nazi occupation in France, he was protected and hidden as a Jew at the Haute Provence Observatory. It is remarquable that the first extrasolar planet 51 Peg b was discovered in this observatory. So there are two commemorative plaques in the Haute Provence Observatory, the first one in memory of the Jews hidden during the war in this observatory and the the second one to celebrate the discovery of 51 Peg b, the first exoplanet.
Gabriel Rémy (1945) wrote in a science popularization book : `*Who knows if we will succeed in some days to detect changes of light emitted by some close stars when an invisible and dark object, like a planet, will periodically cross the field ?*' (Rémy 1945). Rémy was a priest, amateur-astronomer and interested also by microscopic science.
Otto Struve (1897-1963) indicated, in 1952, in the same reference that quoted above: `*...the projected eclipse area is about 1/50th of that of the star, and the loss of light in stellar magnitude is about 0.02.*' (Struve1952). Struve was born in Russia in a family containing several famous astronomers. He made all his career in United States of America where he was a very important astronomer. Among many other subjects of interest and studies, he was very interested by the research of the life in Universe, during a time when only very few astronomers were interested by this research.
These quoted papers or books clearly indicate that the research of planets orbiting other stars than the Sun was a subject . We emphasize that all these papers, even written by professional astronomers, are more often published in science popularization books or magazines and never published in the most famous peer reviewed astronomical journals
Let us finally note that the spectroscopy of transit  as a tool to explore their atmosphère was proposed in 1992, well before the detection of 51 Peg b (Schneider 1994a). Also, the dynamical behavior of circumbinary transiting planets was predicted in 1994 (Schneider 1994b), well before its observational confirmation for Kepler-413(AB) b (Kostov et al. 2014).

4.3 Some 20th century investigations before HD 209458 b

The first exoplanet transit was discovered for a planet already known from radial velocity detection (HD 209458 b). Before that discovery, some systematic searches for transits were made, with two approaches: 1/ search for transits for a few suitably selected stars 2/ systematic search for transits on very large stellar samples.
 The first approach was proposed by Doyle et al. (1984) and Doyle (1985); the idea was to select stars for which the rotation axis, infered from the $i = \arcsin(P_* V\sin i / 2\pi R_*)$ lies in the sky plane, Assuming that the planet orbit is in the star equatorial plane, potential planets should have a large probability of transits.  A special case of an a priori planet favourable orbital orientation was the selection of exlipsing binaries, with the assumption
 that the planet and binary orbits are coplanar. The latter approach was implemented for CM Dra with the Transit of Extrasolar Planets (TEP) network (Deeg et al. 1997). It was the first systematic detection programme of a planetary transit (and by the way, the first
systematic search for circumbinary planets and planets around DM stars). The second approach was based on large stellar samples to compensate the low geometric probability $R_*/a$ of transits (where *a* is the planet orbital radius). The first actual implementation was the FRESIP proposal (Borucki et al. 1996), which was finally lauched as the Kepler mission. It was greatly facilitated with the development of CCDs in the 80's, although it could in principle have been possible much before with a Lallemand electronic camera equipping a wide field telescope (Lallemand et al. 1970). And even, as suggested by Belorizky (1938), it could have detected (by great chance) exoplanet transits on some individual stars decades befors the discovery of the transit of HD 209458 b. Finally, the search for atmospheric composition by transmission spectroscopy was suggested before this latter discovery (Schneider 1992).

**5. Back to the future: $\beta$ Pic**

The exoplanet $\beta$ Pic b was discovered by imaging in 2008 (Lagrange et al 2009). Because of the presence of a circumstellar disk of gas and dust which is oriented edge-on to Earth, the star $\beta$ Pic was considered during several years before the detection of 51 Peg b, as a good candidate for the direct detection of the first exoplanet. So this star has been extensively studied. As early as 1993, some spectroscopic events are interpreted as the signature of the vaporisation of comet-like bodies (Lecavelier des Etangs et al 1993). Moreover a detailed study of previous observations revealed that the star showed some light variations in 1981 which can be possibly interpreted as a planetary transit (Lecavelier des Etangs et al 1995). Another observation of a similar variation is necessary to confirm the exoplanet transit. Studies of the planet $\beta$ Pic b indicate that a transit of this planet in front of its star is possible in 2017-2018. This transit prediction is actually depending of the orbital eccentricity of the planet. In case of a low eccentricity orbit, the expected period is 18 years, and in case of a high eccentricity orbit, the period is 36 years. We have to wait up to around 2018 to obtain an answer and to obtain a confirmation of a transit. The duration of the transit is estimated to approximatively 10 hours. Several observational campaigns are dedicated to accurate ground observations of $\beta$ Pic. A nano-satellite PicSat was designed by astronomers of the Paris-Meudon observatory specially to detect and observe accurately a possible transit on $\beta$ Pic (see for example Nowak et al 2017). This satellite was launched on January 12, 2018. It is also expected to discover comets and study the circumstellar disc inhomogeneities. If the transit is confirmed in the near future, we could obtain better observations of the following transit in 2053, when the period will be known more accurately and when we could use very large and extraordinary outstanding future instruments. So, will Beta Pictoris b win the title of the first detected exoplanet ?

**6. The future is already present: transits and extraterrestrial civilisations**

Nowadays we can notice that the detection of extrasolar planets has led to some changes in astronomical publications. As we said above, the studies predicting the discovery of extrasolar planets from their transits in front of their star, were more often published in books or magazine of popular science. The discovery of exoplanets and the expectation that these studies may allow to detect an extraterrestrial life in a more or less distant future has extended our area of scientific research. Some studies about planet transits, which would considered as pure science fiction until just recently, are now published in astronomical journals with peer review. We give now some examples. In 2005, Luc Arnold published a study about transit light-curve signatures of artificial objects (Arnold 2005). These artificial objects would be built and put into orbit by some civilisations living on extrasolar planets, and willing to make themselves known. These artificial objects would be designed is such a manner their transits are completly different of all natural planet transits as we know them. The cases of natural planets transits include single planets, or with moons, or with rings. These artificial objects could be for example simple objects of unusual shapes, as triangles, or furthermore a fleet of objects. Obviously, that would imply a very high level of civilization for the extraterrestrial creatures imaginating and building some such artificial planets. The Arnold's paper has inspired other studies. As an example, in 2015, Korpela et al. studied how extraterrestrial civilisations would change the transits that we could observe, by illuminating the dark side of their planet (Korpela et al 2015). In 2016, Kipping and Teachey considered the point of view of inhabitants of the planet Earth who would broadcast or cloak their existence to extraterrestrial civilizations (Kipping and Teachey 2016). This would be possible with laser emission. These last three papers were published in journals with peer review, i.e. Astrophysical Journal and Monthly Notices of the Royal Astronomical Society.

Very recently the hypothesis of artificial transits or structures made by extraterrestrial civilizations has been considered to explain some observations of a star (KIC 8462852) still unexplained with purely physical processes. Let us notice that when the first pulsars have been discovered but not yet explained by astrophysical theories, they were called LGM1 and LGM2, as Little Green Man 1 and Little Green Man 2, but it was only something as a private joke, far from being published in a scientific journal. To be complete, we wish recall that of course a scientific explanation was found to explain pulsar observations. We know now that pulsars are neutron stars, the end of the life of some massive stars.

## 7. Conclusion

Observations and studies of planetary transits of planets, or other objects, in front of their star form a very fruitful part of astronomical research. Studies of transits in the solar system began à long time ago and their history is very interesting because containing a lot of unexpected episodes. The first observation of a planet transit in front of the Sun, that is the observation of the Mercury transit carried out scientifically by Pierre Gassendi in 1631, is an important step in the history of astronomy. Nowadays, more than half of the several thousands of extrasolar planets were discovered by the transit method, particularly thanks to the Kepler space telescope, we found several premonitory and visionary studies, published several decades before the discovery of the first exoplanet in 1995. These studies foresaw that extrasolar planets could be detected by precise and continue observations of luminosities of stars. The regular and periodic decrease of luminosities of these stars due to the transits of a planets could be detected by some instruments existing at this time. One can wonder why these premonitory studies were ignored. Actually, they were not published in the most famous refeered journals. Maybe the reason is that they were too different and too new in comparison to the research results published at these times. Maybe as well because the subject was not considered as a serious one. However, if the study of Belorizky (Belorizky 1938) published in a French journal for amateur astronomers was not ignored and if some survey observation programs were carried out, perhaps extrasolar planets would be discovered much earlier than the discovery of 51 Peg b in 1995.

The discovery of exoplanets since the discovery of 51 Peg b, non only opened a new research area very fruiful and successful, but also changed some study methods at it can be seen in astronomical publications. Imagination is being always a part of the scientific research, but occupies now a more significant part, larger than ever. A so rapid evolution of the scientific methods was very rarely observed.

How can we know if these studies which take into account non only the existence but also the intelligence and the intentions of the extraterrestrial civilizations, will be considered in the future as very clever and premonitory, or on the contrary as somewhat naive and a little ridiculous ?

**Acknowledgments**:
We thank Tsevi Mazeh for additional information about the psalm 19, the historians Jean-Patrice Boudet and Michel-Pierre Lerner for passionate and very helpful discussions, Guy Artzner for indicating the paper of Belorizky, Mira Véron for giving information about this astronomer, Jean-Yves Giot for indicating the book of Rémy, and Szilard Csizmadia for suggesting the study of Algol